\documentclass[aps,prl,twocolumn,amssymb,superscriptaddress]{revtex4}

\usepackage{graphicx}

\begin{document}

{\bf \noindent Comment on ``Density of States and Critical Behavior of 
the Coulomb Glass''}

\vspace{0.2cm}

In a recent Letter \cite{Surer}, Surer {\it et al.}\ concluded that
their simulation results are consistent with the Efros Shklovskii 
prediction for the density of states in the three-dimensional Coulomb 
glass. Here, we show that this statement has  no relevance concerning 
the problem of the asymptotic behavior in the Coulomb gap since it is 
based on unjustified assumptions. Moreover, for the random-displacement 
Coulomb glass model, we demonstrate that a part of the density of states 
data by Surer {\it et al.}\ erroneously exhibit a broad gap. This is 
related to the staggered occupation being instable contrary to 
\cite{Surer}.

In detail, Efros and Shklovskii considered systems of localized charges
(see the review \cite{ESrev}). They showed that the single-particle 
density of states $\rho$ vanishes when the energy $E$ approaches the 
chemical potential, which equals 0 here. For stability with respect to 
arbitrary single-particle hops, they analytically derived that, as 
$E \rightarrow 0$, $\rho(E) = a |E|^\delta$ with $\delta = 2$ in the 
three-dimensional case. 

Surer {\it et al.}\ \cite{Surer} state correctly, that the literature 
situation concerning quantitative numerical studies of this Coulomb gap 
is not yet satisfactory. Then they claim that their $\rho(E)$ data ``can 
be fit very well with a form $\propto |E|^\delta$'' with $\delta$ 
close to 2. This is puzzling for their samples are considerably smaller 
than the samples in our old work \cite{MRD}, where clear deviations from 
$\propto E^2$ were found. 

Inspection of Fig.\ 1 of \cite{Surer} uncovers a first problem: The 
simulation data seem to be approximated by the ansatz 
$\rho(E) = a |E|^\delta + b$ instead of by the Efros Shklovskii power 
law, but this is not mentioned in the text of \cite{Surer}. 

For a detailed analysis, we digitized Fig.\ 1 of \cite{Surer} by means 
of WinDig. The data for the disorder strength $W = 0.4$ are re-plotted 
in a log-log representation in Fig.\ 1(a) here. Their precision is 
confirmed by a fit to $\rho(E) = a |E|^\delta + b$ for $|E| < 0.3$ and
$L = 14$ yielding $\delta = 1.84(3)$, in almost perfect agreement 
with \cite{Surer}.

Figure 1(a) shows that these $\rho(E)$ data clearly do not follow a pure 
power law, so that the quality of the fit could be ensured only by the
unphysical constant $b$. 

One might argue, $b$ emulates the finite-size effects not analyzed in 
\cite{Surer}, being important due to the long range of the interaction 
\cite{MRD}. However, the data of \cite{Surer} exhibit an unexpected 
feature contradicting this idea: For sample-edge length $L = 14$, $b$ is 
clearly larger than for $L = 8$, although $b$ should vanish as 
$L \rightarrow \infty$ \cite{GP}. A second con arises from the $\rho(E)$ 
by Surer {\it et al.}\  being far larger than the data from Fig.\ 3(b) 
in \cite{MRD} for small $|E|$, albeit another random potential was used 
there (compare \cite{GP}).

To clarify these points, we performed own simulations for $W = 0.4$ and 
various $L$ using periodic boundary conditions and minimum image 
convention. The results are included in Fig.\ 1(a). For large $|E|$, 
our $\rho(E)$ obtained by relaxation concerning all single-particle hops 
nicely agree with $\rho(E)$ from \cite{Surer}, but for $|E| < 0.3$,
within the whole fit region of \cite{Surer}, they are considerably 
steeper and smaller. 

According to the original reasoning of the Coulomb gap \cite{ESrev},
this difference may arise from the minimum search in \cite{Surer} not 
considering all single-particle hops. Figure 1(a) includes two $\rho(E)$ 
curves which we obtained by relaxation only via short hops ranging up to 
third-next neighbors (length $\le \sqrt 3$). These relations 
support our hypothesis.

The situation is even more inconsistent in case of the 
random-displacement model: Compared to \cite{Surer}, we found
far stronger broadening of the $\delta$ peaks of
a charge-ordered (NaCl) occupation of the lattice. Figure 1(b) shows 
that already for $W = 0.2$, a weak Gaussian random displacement,
broadening causes $\rho(0) \approx 0.11$. Hence, a Coulomb gap is 
formed by relaxation, in which the perfect charge order claimed in Fig.\ 
2 of \cite{Surer} for $T \rightarrow 0$ is disturbed.

\begin{figure}[t]
\includegraphics[width=0.84\linewidth]{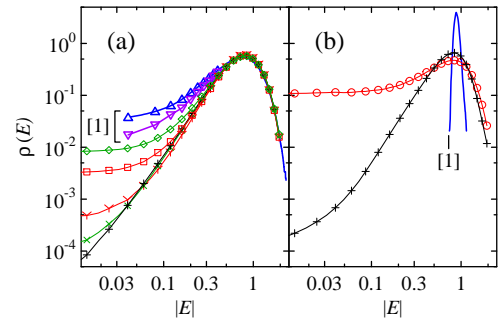}
\caption{(Color online) $\rho(E)$ for Gaussian random-potential (a) and 
random-displacement (b) Coulomb glass models with $W = 0.4$ and 0.2, 
respectively. (a) From \cite{Surer}: $\bigtriangledown$, $L = 8$, and 
$\bigtriangleup$, $L = 14$. Own: {\sf Y}, $L = 6$, $\times$, $L = 20$, 
and $+$, $L = 70$, for complete relaxation via single-particle hops, 
whereas $\square$, $L = 6$, and $\Diamond$, $L = 20$, for relaxation 
only via short-range hops (see text). (b) From \cite{Surer}: continuous 
line. Own: $\bigcirc$, charged-ordered occupation without relaxation, 
$+$, relaxed concerning single-particle hops. All three for 
$L = 14$. Error bars of own data are omitted since they do not exceed 
the symbol size.}
\label{fig}
\end{figure}

\vspace{0.2cm}

\noindent
A.\ M\"obius and M.\ Richter\\
\indent {\small 
IFW Dresden, POB 27 01 16, D-01171 Dresden, Germany}
\indent {\small 
e-mail: a.moebius@ifw-dresden.de}

\vspace{0.2cm}

\noindent
Received {\today}

\noindent
DOI:


\end{document}